\pdfoutput=1
\documentclass[aps,twocolumn,showpacs,preprintnumbers,amsmath,amssymb,footinbib]{revtex4}
\usepackage[colorlinks=true, pdfstartview=FitV, linkcolor=red, citecolor=blue, urlcolor=blue,pdfsubject={}, pdfkeywords={Heavy ion collisions,quark gluon plasma}]{hyperref}
\usepackage{graphicx}
\newcommand{\arXiv}[1]{\href{http://arXiv.org/abs/#1}{[arXiv:#1]}}
\newcommand{\beq}{\begin{equation}}
\newcommand{\eeq}{\end{equation}}

\def\arnps#1#2#3{Ann.\ Rev.\ Nucl.\ Part.\ Sci.\  {\bf #1}, #2 (#3)}
\def\atmp#1#2#3{Adv. Theor. Math. Phys. {\bf #1}, #2 (#3)}

\def\epja#1#2#3{Eur.\ Phys.\ Jour.\ A{\bf #1}, #2 (#3)}

\def\epja#1#2#3{Eur.\ Phys.\ J. A {\bf #1}, #2 (#3)}
\def\ibid#1#2#3{{\it ibid.} {\bf #1}, #2 (#3)}

\def\jhep#1#2#3{Jour. High Energy Phys. {\bf #1}, #2 (#3)}

\def\jpg#1#2#3{Jour. Phys. G {\bf #1}, #2 (#3)}

\def\npa#1#2#3{Nucl. Phys. A {\bf #1}, #2 (#3)}
\def\npb#1#2#3{Nucl. Phys. B {\bf #1}, #2 (#3)}

\def\plb#1#2#3{Phys. Lett. B {\bf #1}, #2 (#3)}
\def\pos#1#2#3{PoS {\bf #1}, #2 (#3)}
\def\prc#1#2#3{Phys. Rev. C {\bf #1}, #2 (#3)}
\def\prd#1#2#3{Phys. Rev. D {\bf #1}, #2 (#3)}
\def\prl#1#2#3{Phys. Rev. Lett. {\bf #1}, #2 (#3)}

\def\ppnp#1#2#3{Prog. Part. Nucl. Phys. {\bf #1}, #2 (#3)}
\def\ptps#1#2#3{Prog. Theor. Phys. Suppl. {\bf #1}, #2 (#3)}

\def\zpc#1#2#3{Z. Phys. C {\bf #1}, #2 (#3)}
\preprint{BNL-94342-2010-JA, KUNS-2310, NIKHEF 2010-040, RBRC-871}
\begin{document}
\title{How Wide is the Transition to Deconfinement?}
\author{Adrian Dumitru$^{a,b}$, Yun Guo$^{c,d}$, 
Yoshimasa Hidaka$^{e}$, Christiaan P. Korthals Altes$^{f,g}$, 
and Robert D. Pisarski$^h$}
\affiliation{
$^a$Department of Natural Sciences, Baruch College,
17 Lexington Avenue, New York, NY 10010, USA\\
$^b$RIKEN/BNL Research Center, Brookhaven National Laboratory, 
Upton, NY 11973, USA\\
$^c$Department of Physics, Brandon University,
Brandon, Manitoba, R7A 6A9 Canada\\
$^d$Winnipeg Institute for Theoretical Physics, Winnipeg, Manitoba, Canada\\
$^e$Department of Physics, Kyoto University, Sakyo-ku, Kyoto 606-8502, Japan\\
$^f$Centre Physique Th\'eorique au CNRS Case 907,
 Campus de Luminy F-13288 Marseille, France\\
$^g$Nikhef Theory Group, Science Park 105, 1098 XG Amsterdam, The Netherlands\\
$^h$Department of Physics, Brookhaven National Laboratory, 
Upton, NY 11973\\
}
\date{\today}

\begin{abstract}
Pure $SU(3)$ glue theories exhibit a deconfining phase transition at a
nonzero temperature, $T_c$.  Using lattice measurements of the pressure, 
we develop a simple matrix model to describe the transition region,
when $T \geq T_c$.  
This model, which involves three parameters, is used to compute the
behavior of the 't Hooft loop.
There is a Higgs phase in this region,
where off diagonal color modes are heavy, and diagonal modes are light.
Lattice measurements of the latter
suggests that the transition region is narrow, extending only
to about $\sim 1.2 \, T_c$.
This is in stark contrast to lattice measurements of
the renormalized Polyakov loop, which indicates
a much wider width.  The possible implications for the
differences in heavy ion collisions between
RHIC and the LHC are discussed.
\end{abstract}

\pacs{12.38.Mh,12.38.Gc,25.75.Nq}
\maketitle

Heavy ion collisions at the Relativistic Heavy Ion Collider (RHIC) 
have demonstrated a rich variety of unexpected behavior
\cite{hydro}.  Notably, in peripheral collisions the elliptical flow can
only be described by nearly ideal hydrodynamics, with a very small
ratio between the shear viscosity, $\eta$, and the entropy density, $s$.
The differences
between collisions at RHIC, and those which
will soon be observed soon at the Large Hadron Collider (LHC), 
will be especially interesting: does the nearly ideal hydrodynamic
behavior, observed at RHIC, persist at the much
higher energies of the LHC?

One approach to deconfinement exploits the analogy to
${\cal N} = 4$ supersymmetric gauge theories:
using the AdS/CFT correspondence, 
such theories are computable analytically 
in the limit of infinite coupling, for
an infinite number of colors \cite{ads_reviews}.
By introducing a potential for the dilation field, the
behavior of the entropy density near the deconfining phase transition,
at a temperature $T_c$, can be fit from measurements
on the lattice \cite{gubser,kiritsis,other,gyulassy}.
While the entropy density, $s$, decreases strongly as $T \rightarrow T_c^+$
because it is 
related to Hawking radiation, in AdS/CFT models
the ratio $\eta/s$ remains
completely independent of temperature.  This suggests that
like RHIC, that collisions at the LHC should also be
described by nearly ideal hydrodynamics; see, also, 
Ref. \cite{shuryak}.

In this work we consider a very different approach to the deconfining
phase transition.  It assumes that the coupling is moderate even down
to the transition temperature, $T_c$
\cite{laine}.  We use an elementary matrix model, involving three
parameters, to parametrize the behavior of the deconfining phase transition.
A version of this model with one parameter was first proposed by
Meisinger, Miller, and Ogilvie
\cite{ogilvie1}.  Similar models arise for theories in which one (or more)
spatial directions are of femtoscale size
\cite{ogilvie2,ogilvie3,matrix_other,aharony}.

The parameters of the model are fixed
from lattice measurements of
the pressure \cite{pressure2,pressure3,panero,pressure_dyn}.
It then predicts how the 't Hooft loop 
\cite{interface1,interface2,interface3,interface4,interface5} changes
with temperature near $T_c$, which we compare to the results of
lattice simulations \cite{interface_lattice1,interface_lattice2}.  
Further, the model predicts that for a range of temperatures
above $T_c$, there is a Higgs
phase, where correlation functions of electric fields are a mixture
of heavy and light modes, from fields which are off diagonal, and diagonal,
in color, respectively.  This may help to understand the 
results of lattice simulations 
\cite{pressure3,gauge_inv_mass_su3,gauge_inv_mass_su2,gluon_mass_1}, which are
otherwise somewhat puzzling.

The most direct prediction of our model is for the 
expectation value of the Polyakov loop.  
For the pure glue $SU(3)$ theory, lattice simulations find that 
the (renormalized) 
Polyakov loop vanishes below $T_c$, jumps to $\sim 0.4$ at
$T_c^+$, and then rises with $T$, until it
is approximately constant above $\sim 4.0 \, T_c$ 
\cite{ren_loop1,ren_loop2,ren_loop3}.  
This represents confinement below $T_c$, a complete 
Quark Gluon Plasma (QGP) at high temperature, and a 
``semi''-QGP in between \cite{semi1,semi2,higgs,yk}.  
Physically, there is no ionization of color in the confined phase,
total ionization in the complete QGP, and only partial ionization
in the semi-QGP \cite{yk}.
(While we discuss a purely gluonic plasma, we adopt
the common term QGP.)

The principal thrust of this paper is that from 
indirect measurements on the lattice, we suggest
that the width of the semi-QGP is {\it much} narrower than indicated by
present results for the renormalized Polyakov loop:
not up to $\sim 4.0\, T_c$, but only $\sim 1.2 \, T_c$.
We do not understand this discrepancy in detail, 
but suggest a possible reason later.
This discrepancy is the reason why, having fit
the parameters of our model from the pressure, 
we compute both the 't Hooft loop and gluon masses.

While we treat the pure glue theory, our model can be 
extended to QCD, with dynamical quarks \cite{pressure_dyn}.
It is reasonable to assume that in 
QCD, the semi-QGP is like that of the pure glue theory, relatively narrow.
We thus conclude by discussing 
the possible phenomenological implications 
of our results for heavy ion collisions.

How confinement arises in our model can be understood by analogy.  
For a bosonic field, with energy $E$ and chemical potential $\mu$,
the Bose-Einstein statistical distribution function is 
\beq
n(E,q) = \frac{1}{{\rm e}^{(E-\mu)/T} - 1}
= \frac{1}{{\rm e}^{E/T - 2 \pi i q} - 1} \; .
\label{bose_ein}
\eeq
Instead of taking $\mu$ to be real, as in ordinary thermodynamical systems,
for the purposes of the analogy we take it to be purely
imaginary, and define $\mu = 2 \pi T \, i \, q$, where $q$ is real.

Doing so, $n(E,q)$ is clearly a periodic function of $q$,
invariant under $q \rightarrow q + 1$.  Thus we can choose to define
$q$ to lie within the range from $-\frac{1}{2}$ to $+\frac{1}{2}$.  

Now {\it assume} that we integrate
over $q$, with a distribution which is flat in $q$. 
Expand for large energy, so that the first term is the Boltzmann statistical
distribution function.  Given the assumed distribution in $q$,
the integral of this term vanishes,
\beq
{\rm e}^{- E/T}\; \int^{+1/2}_{-1/2} \; {\rm e}^{2 \pi i q} \; dq = 0 \; .
\label{integral_q}
\eeq
Indeed, we can expand the Bose-Einstein distribution function term by term
in powers of Boltzmann factors,
${\rm e}^{-E/T + 2 \pi i q}$ \cite{aharony}; 
doing so, the integral over each and every term obviously vanishes.  The same
is true for the Fermi-Dirac distribution function as well.

Thus a flat distribution in $q$ represents the confined phase.  To
represent a phase with partial deconfinement, one integrates over a
limited region, say $q:-q_0 \rightarrow + q_0$, with $q_0 < \frac{1}{2}$.
Complete deconfinement occurs when one integrates over a distribution
which is a delta-function in $q$.

This example appears somewhat artificial.  For a given $q$,
the statistical distribution functions are
complex valued, and so only integrals over $q$ can possibly represent
physical quantities.  Indeed, the grand canonical ensemble is characterized
by a fixed value for the chemical potential, and not by an integral
over $\mu$'s.  

Nevertheless, precisely this mechanism arises 
for the deconfining phase transition in a $SU(N)$ gauge theory.
Consider the expansion about a background
field for the time-like component of the vector potential,
\beq
\left(A^{cl}_0\right)_{a b}  
= \frac{2 \pi T }{g} \; q_a \; \delta^{a b} \; ;
\label{background}
\eeq
$a$ and $b$ are colors indices, running from $1\ldots N$.
For nonzero $q_a$'s, this background field acts like an
imaginary chemical potential for the diagonal elements of the gauge group.
Integration over the $q_a$'s arises from imposing Gauss' law
for those elements of the gauge group 
\cite{interface2}.

This background field generates a non-trivial
expectation for the Polyakov loop, $\ell$, which is the color
trace of the thermal Wilson line, ${\bf L}$:
\beq
\ell = \frac{1}{N} \; {\rm tr}\, {\bf L} \;\; ; \;\;
{\bf L} = {\cal P} \exp\left( i g \int^{1/T}_0 A_0 \; d \tau \right) \; ;
\label{loop}
\eeq
$\cal P$ represents path ordering, 
$T$ is the temperature, and $\tau$ the imaginary time, 
$\tau: 0 \rightarrow 1/T$.

The lattice demonstrates that near $T_c$, the expectation value of the
Polyakov loop is not near one, and decreases as the temperature does.
In such a region, the eigenvalues 
of the logarithm of the thermal Wilson line are nonzero.
Taking an ansatz such as Eq. (\ref{background}) is the simplest
possible way to model this.  We do not attempt to derive the distribution of
these eigenvalues, but to guess that from lattice results.

Since the gauge potential $A_0$ is an element of 
$SU(N)$, $\sum_{a=1}^{N} q_a = 0$, modulo one,
there are $N-1$ independent $q_a$'s.
At infinite $N$, the $q_a$'s form
a continuum, and the example of Eq. (\ref{bose_ein}) is exact; see,
{\it e.g.}, computations on a femtosphere at $N=\infty$ \cite{aharony}.
For two colors, we can choose
the eigenvalues to be $q_1 = - q_2$; for three, $q_1 = - q_2$, and $q_3 = 0$.

In the presence of the background field of Eq. (\ref{background}),
a potential for the $q_a$'s is generated at one loop order 
\cite{interface1,interface2,interface3,interface4,interface5},
\beq
{\cal V}_{pt}(q_a)  \; = \;
\frac{2 \pi^2 T^4}{3} \;\sum_{a,b = 1}^N
\; q_{ab}^2 \left( 1 - |q_{ab}| \right)^2  
- \; (N^2-1) \frac{\pi^2 T^4}{45} \; .
\label{pert_pot}
\eeq
where $q_{a b} = q_a - q_b$, defined modulo one.  
The minimum is at $q_a = 0$, where
$- {\cal V}_{pt}(0)$ is the pressure
for an ideal gas of gluons.

The potential ${\cal V}_{pt}(q_a)$ enters in computations
of the 't Hooft loop.
It is useful to consider deconfinement as a type of spin system.
A pure $SU(N)$ gauge theory
has $N$ degenerate vacua, where the thermal Wilson line $\bf L$
equals one of the $N$ roots of unity, 
\beq
{\bf L} = {\rm e}^{2 \pi i j/N}\; {\bf 1} \; ,
\eeq
$j=0\ldots (N-1)$.  The usual vacuum, with $j=0$ and
${\bf L} ={\bf 1}$, corresponds to all $q_a = 0$.  
A $Z(N)$ vacua with $j = 1$ and
${\bf L} = {\rm e}^{2 \pi i/N} {\bf 1}$ 
corresponds to $N-1$ $q_a$'s $=1/N$, and the
remaining element $= -1 + 1/N$.

At high temperature in the complete QGP,
the theory lies in one spin state, which we can
choose to be $j=0$.  One can compute tunneling between two degenerate vacua 
by constructing a box which is long in one spatial direction, with
$j=0$ at one end, and $j=1$ at the other.  An interface between
the two ordered states forms in the center of the box, with the interface
tension between the two computable semi-classically,
using the potential ${\cal V}_{pt}(q_{a})$ 
\cite{interface1,interface2,interface3,interface4,interface5}.  
This interface is equivalent to a 
't Hooft loop which wraps around the center of the box 
\cite{interface3}.

As the temperature decreases and $T$ approaches $T_c$, 
domains with $j \neq 0$ form and grow in size.  They become
increasingly probable, until at $T_c^-$ and below, as a spin
system the vacuum is completely disordered, a sum over many spin domains.

We want to add terms to the effective potential which model the transition
to deconfinement.  We could add 
perturbative corrections to ${\cal V}_{pt}(q_{a})$, which
have been computed to $\sim g^3$ 
\cite{interface4}, 
but invariably they give $q_a = 0$ (or a $Z(N)$ equivalent state)
as the vacuum.  With a complete theory, 
such as the monopole model of Liao and Shuryak \cite{shuryak},
this potential could be computed directly.

We adopt a more modest approach, attempting to guess the
form of the non-perturbative potential.
We fit the coefficients which enter to lattice results for the
pressure, and then use it to compute other quantities.  
The advantage of our approach is that we can 
compute quantities not just in, but near thermal equilibrium.
Such quantities, like the shear viscosity \cite{yk}, are much harder
to extract on the lattice.

Since the Polyakov loop is an order parameter for deconfinement, a natural
guess is that the non-perturbative potential involves $Z(N)$ invariant
elements of the Lie group.  The first such term is the adjoint 
loop \cite{matrix_other,semi1,semi2,higgs,yk,lattice_eff,pnjl}.  
Instead, following the authors
of Ref. \cite{ogilvie1}, and computations
of the 't Hooft loop 
\cite{interface1,interface2,interface3,interface4,interface5}, we 
write a potential which is a polynomial in the $q_a$'s.  

There are several symmetries which any potential of the
$q_a$'s must satisfy.  
It must be periodic in each $q_a$, with
$q_a \rightarrow q_a + 1$.  It must also be invariant under $Z(N)$
transformations, where $N-1$ of the $q_a$'s shift by $1/N$, and the
last element, by $-1 + 1/N$.   Lastly, if we interchange the ordering
of the $q_a$'s, we can change $q_{a b} \rightarrow q_{b a} = - q_{a b}$.
These symmetries can be satisfied by constructing a potential as a
function of $q_{a b}(1 - q_{a b})$.

We can still form an infinite number of terms by tying together the
color indices in different ways; see, {\it e.g.}, the examples
at two \cite{interface4} and three \cite{aharony} loop order.
We adopt the simplest approach, and take terms
like those which arise at one loop order, Eq. (\ref{pert_pot}),
which involve a sum over one $q_{a b}$:
$$
{\cal V}_{non}(q_a) = T^2 \; T_c^2 \; \sum_{a,b = 1}^N \left(
c_1 \; |q_{ab}|( 1 - |q_{ab}|) \right. 
$$
\beq
\left. 
+ c_2 \; q_{ab}^2 \left( 1 - |q_{ab}| \right)^2  + c_3 \right) \; .
\label{non}
\eeq
The model of Ref. \cite{ogilvie1} involves terms
$\sim c_1$ and $c_3$; these are kept in fixed ratio, given
by the second Bernoulli polynomial.   Instead, we allow
$c_1$ and $c_3$ to vary independently.  This helps to avoid a
pathology of the model of Ref. (\cite{ogilvie1}), where the pressure
is negative at $T_c$.

We also introduce
a term $\sim c_2 \, q_{a b}^2 (1-q_{a b})^2$; this is 
proportional to the perturbative term in Eq. (\ref{pert_pot}), 
and is related to the fourth Bernoulli polynomial.

We take all of the non-perturbative terms to be
$\sim T^2$, since lattice simulations indicate that in the pure glue theory,
the leading corrections to terms $\sim T^4$ are $\sim T^2$ 
\cite{ogilvie1,ogilvie2,higgs,fuzzy}.  
There is obviously
no fundamental reason why other terms, such as those
$\sim 1$, could not also appear.  

When the $q_a$'s develop an expectation value, this
represents symmetry breaking for an adjoint scalar field, $A_0$,
coupled to an $SU(N)$ gauge field, the $A_i$'s
\cite{higgs}.  
As an adjoint scalar, though, there is no 
strict order parameter which distinguishes between the symmetric and
broken phases.  Thus there need not be a phase transition
in going from the symmetric phase, the complete QGP, to the
``broken'' phase, the semi-QGP.

If there were such a phase transition, it
would represent a second transition, 
above $T_c$, separate from that for deconfinement.
While possible, in a pure $SU(N)$ gauge
theory lattice simulations only find evidence for one phase transition,
at $T_c$ \cite{pressure2,pressure3,panero,pressure_dyn}.  
To avoid a phase transition between the complete and semi-QGP,
it is essential that the non-perturbative potential
has a term which is linear in the $q_a$'s.
Assume that the effective potential only involved 
terms such as $\sim q_{a b}^n (1-q_{a b})^n$ for $n \geq 2$.
For small $q_a$, these are of quadratic or higher order in the $q_a$'s,
and of necessity, there would then be a phase transition when the
$q_a$'s developed a nonzero expectation value.  This
transition might be of either first or second
order, but there would be a phase transition.  
When $c_1 \neq 0$, though, a term
{\it linear} in the $q_a$'s ensures
that there is no such phase transition.
Instead, even for high temperature, there is always a small
but non-zero expectation value
for the $q_a$'s, $<q_a> \, \sim 1/T^2$;
that is, the theory is always
in a Higgsed phase.  As we shall see however, 
this point is somewhat academic.  For the parameters
relevant to two and three colors, the
region in which Higgsing matters is {\it very} narrow.

We remark that effective theories on the lattice often
exhibit phases with broken symmetry \cite{lattice_higgs}.
The necessity of such a broken phase near $T_c$ does not
seem to have been appreciated previously, though.

\begin{figure}[<as soon as possible>]
\includegraphics[width=60mm]{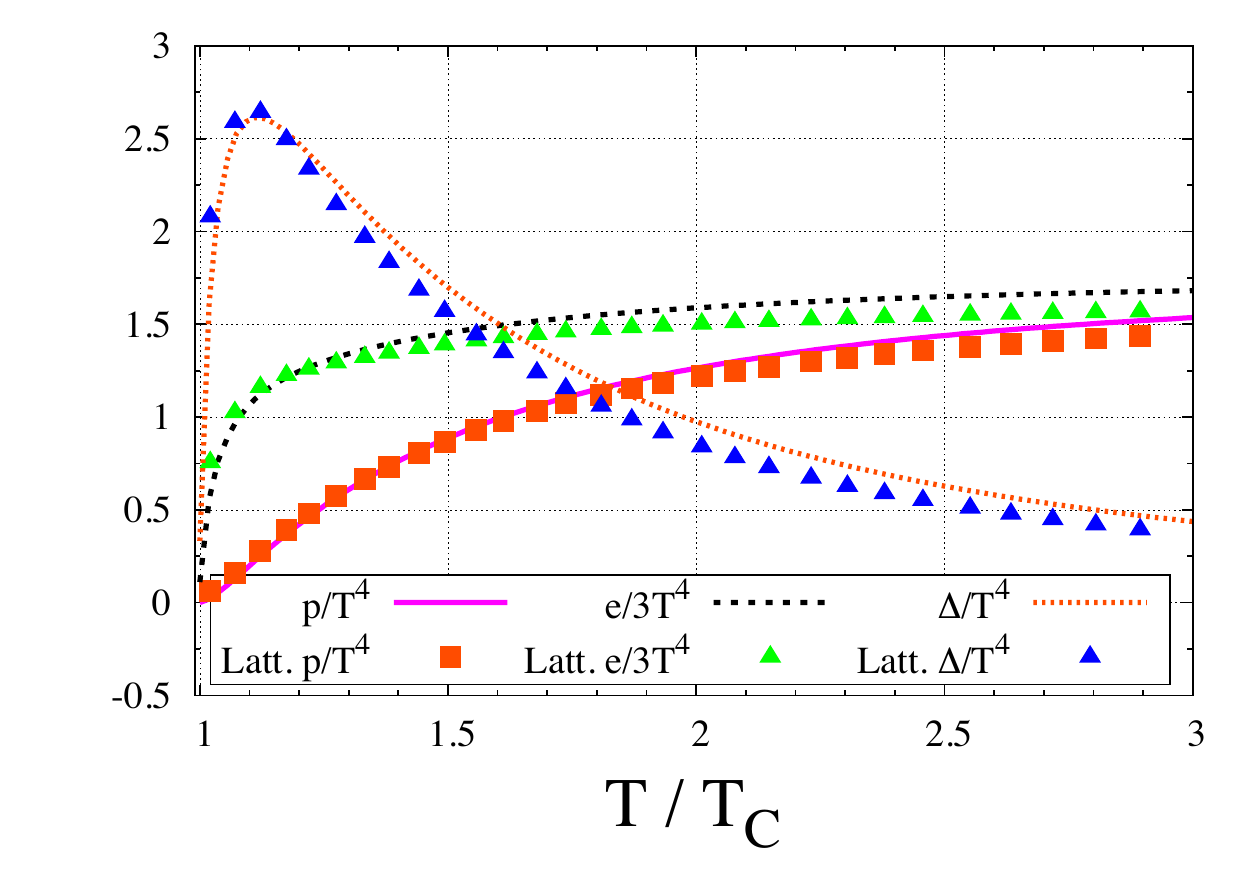}
\caption{Comparison of lattice results for $SU(3)$ pure gauge to the model, for the pressure, energy density, and interaction measure.}
\label{fig:3pressure}
\end{figure}

\begin{figure}[<as soon as possible>]
\begin{center}
\includegraphics[width=60mm]{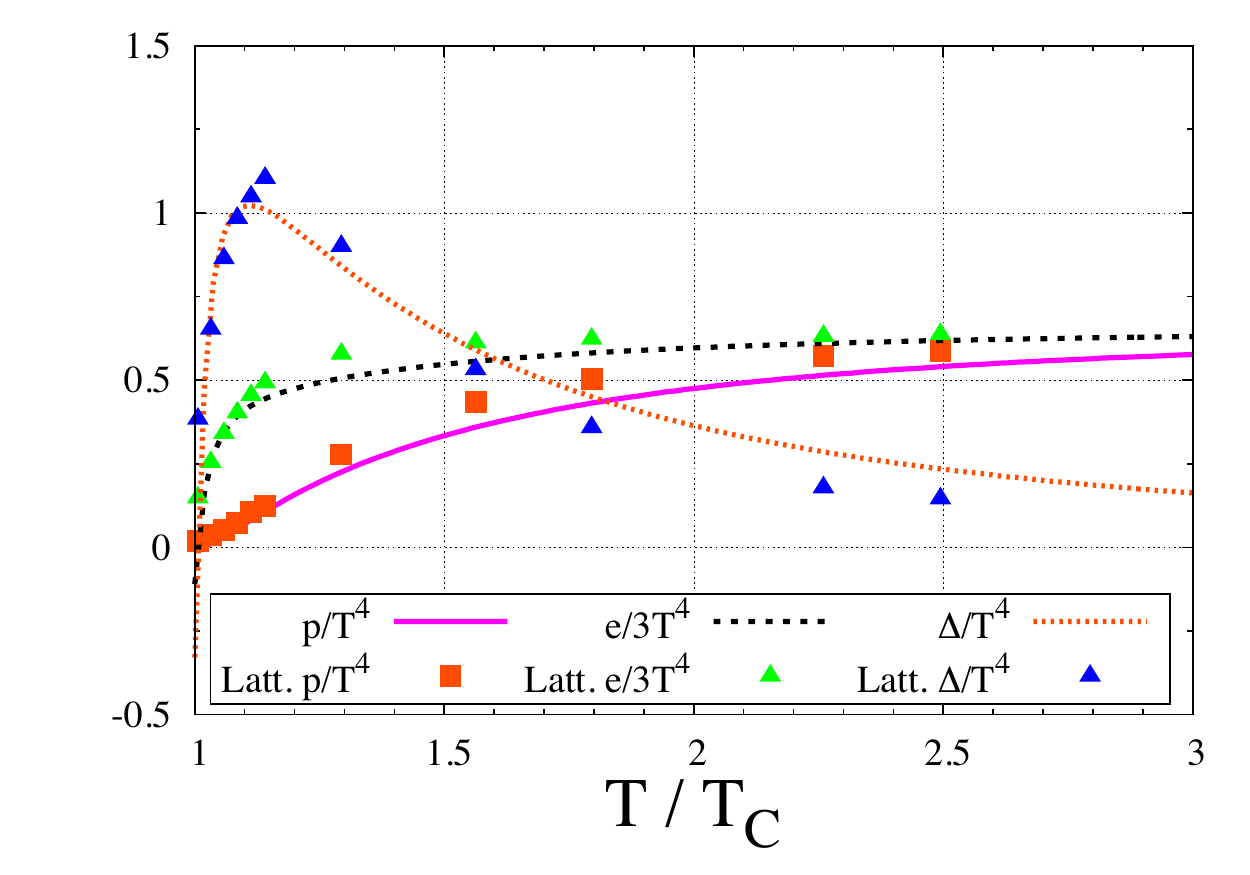}
\caption{Comparison of lattice results for $SU(2)$ pure gauge theory to the model, for the pressure, energy density, and interaction measure.}
\label{fig:2pressure}
\end{center}
\end{figure}

To determine the parameters of the model we compare to lattice measurements
of the pressure.  
For three colors, this is illustrated in Fig. (\ref{fig:3pressure});
for two colors, in Fig. (\ref{fig:2pressure}).  If $p(T)$ is the pressure,
and $e(T)$ the energy density, then a more sensitive test of the fit
is also to plot the interaction measure, 
$\Delta= e - 3p$.  Thus in each
figure we plot $p/T^4$, $e/T^4$, and $\Delta/T^4$, both from the lattice,
from Ref. (\cite{pressure2}) for two colors,
and from Ref. (\cite{pressure3}) for three colors. 

The parameters of the fit are
\beq
c_1 = - \, .41488 \, ; \,
c_2 = -\,5.45957\, ; \,
c_3 = 0.21954 \, .
\label{parameters_three_colors}
\eeq
for three colors, and
\beq
c_1 = -\, 0.30267 \, ; \,
c_2 = - \, 5.97440 \, ; \,
c_3 = 0.18341 \, .
\label{parameters_two_colors}
\eeq
for two colors.  

While our model appears to involve three parameters, this is misleading.
One parameter fixes the critical temperature, $T_c$.
A second is chosen so that the pressure vanishes at $T_c$.  
Thus we really have only one free parameter, which is tuned to
fit the behavior of the pressure near $T_c$.

For two colors, our model exhibits unphysical behavior,
as the energy density is negative 
below $\sim 1 \%$ of $T_c$.  This might be corrected by
adding further terms in the nonperturbative potential, such
as higher Bernoulli polynomials.

In any case, since we fix the pressure to vanish at $T_c$, within our
approximations the confined phase has vanishing pressure.  
How to match to a more realistic description of the confined phase is
an important problem, which we defer for now.

Given the effective Lagrangian, it is then straightforward to compute
the 't Hooft loop.  In the complete QGP, the potential includes only
the perturbative potential, ${\cal V}_{pt}(q_a)$, Eq. (\ref{pert_pot});
in the semi-QGP, it is a sum of this and the non-perturbative
potential, ${\cal V}_{non}(q_a)$.

For two colors, as $q_2 = -q_1$ there is only one independent direction, 
and it is direct to compute the tunneling path, and its associated
action, analytically.  The result for the 't Hooft loop is
\beq
\sigma(T) =
\frac{4 \pi^2 T^2}{3 \sqrt{6 g^2(T)}} \; \xi(g^2) \; 
\frac{(1 - (T_c/T)^2 )^{3/2} }{1 - 0.908\, (T_c/T)^2} \; ,
\label{two_color_interface}
\eeq
where
$$
\xi(g^2) = 1 - 0.16459 \, g^2(T) \; .
$$
The factors involving $T_c/T$ are special to the semi-QGP, so that
as $T \gg T_c$, the result reduces to that in the complete QGP 
\cite{interface1}.  The function $\xi(g^2)$ is the correction
$\sim g^2$ in the complete QGP; in plotting, we take 
$g^2(2 \pi T)$ \cite{laine}.  

The 't Hooft loop vanishes
at $T_c$, as expected for a second order phase transition.
From universality, the result in the Ising model is
$\sigma(T) \sim (T - T_c)^{2 \nu}$, with $2 \nu \sim 1.26$; lattice
results in a gauge theory \cite{interface_lattice1} find $2 \nu \sim 1.32$ 
\cite{interface_lattice1}.  Our result, $2 \nu = 1.5$, is
not too far off, as expected for a mean field theory.
We note, however, that because of the term in the
denominator, that the numerical agreement isn't close.  This is 
presumably related to the unphysical behavior of the energy density
near $T_c$, mentioned previously. 

\begin{figure}[<as soon as possible>]
\begin{center}
\includegraphics[width=60mm]{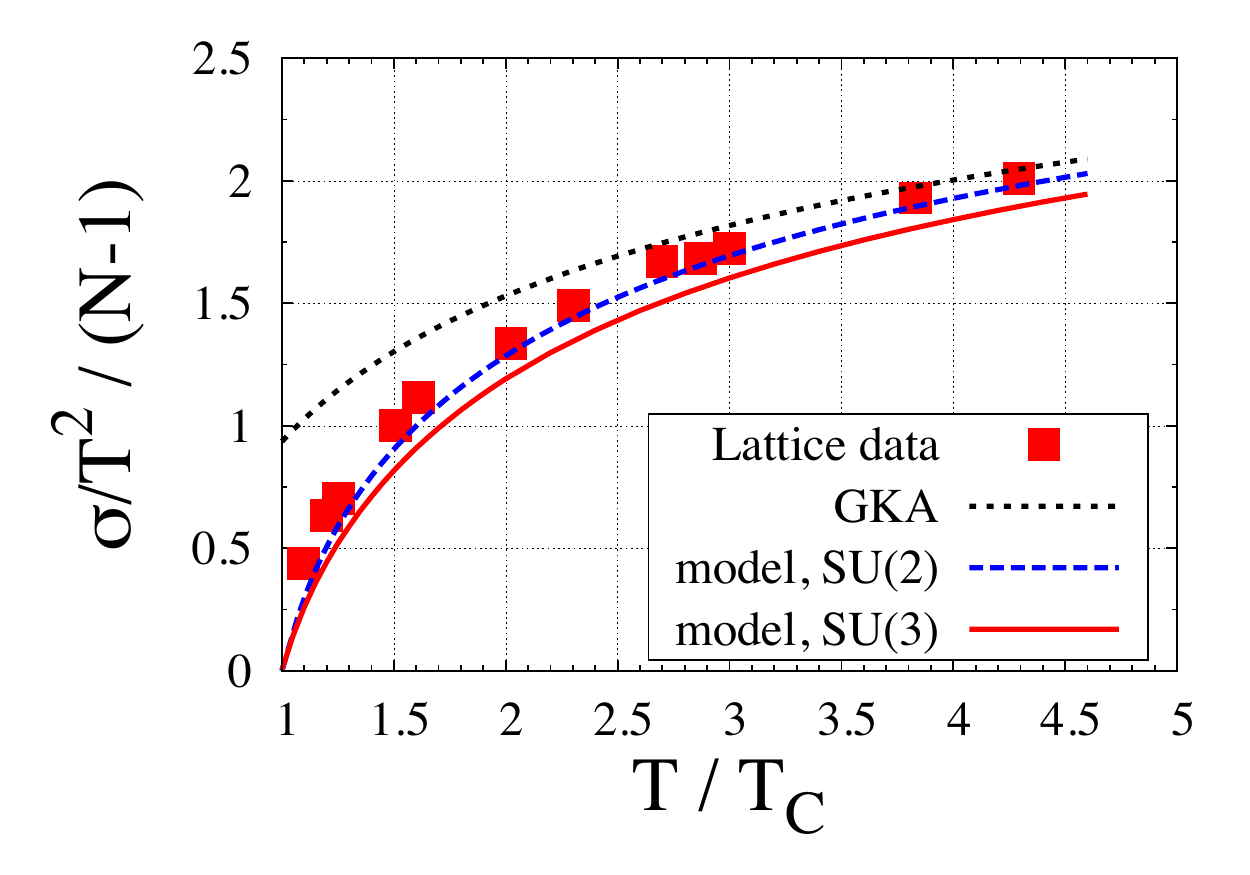}
\caption{The 't Hooft loop for $SU(3)$ pure gauge theory:
lattice data from Ref. \cite{interface_lattice2}, and
GKA, Giovannangeli and Korthals Altes, Ref. \cite{interface4},
the semi-classical computation in
the complete QGP, including corrections of $\sim g^2$.
In our model we show results for two and three colors, 
assuming that the corrections of $\sim g^2$
are identical in the complete and semi-QGP; see text.}
\label{fig:Interface}
\end{center}
\end{figure}

For three colors, in the semi-QGP
the vacua is along $\lambda_3$
(using the Gell-Mann notation), while the path for the 't Hooft loop
depends upon a change in $\lambda_8$.  
The path was determined numerically,
and lies along both $\lambda_3$ and $\lambda_8$.  The action of
the tunneling path was also determined numerically,
and the result for the 't Hooft loop for three colors is 
illustrated in Fig. (\ref{fig:Interface}).
(For $N=2$, we take
$T_c/\Lambda_{\bar{MS}}=1.31$; for $N=3$, $1.14$.
For the same value of 
$T_c/\Lambda_{\bar{MS}}=1.31$, the results unexpectedly coincide.)

Including $\xi(g^2)$,
the semi-classical computation of the 't Hooft loop
in the complete QGP agrees with
lattice simulations above $\sim 4.0 \, T_c$;
below that temperature, they agree with the result in
the semi-QGP 
\cite{interface_lattice2}.  To obtain agreement, however,
it is necessary to include the correction $\xi(g^2)$;
this is computed in the complete QGP, which is
incorrect.  Two things are required to 
compute  $\xi(g^2)$ in the semi-QGP.
First, the potential for constant $q_a$ needs to be computed to two loop
order, expanding about the full potential,
${\cal V}_{pt}(q_a) + {\cal V}_{non}(q_a)$.
Second, corrections to one loop order need to be computed
for the kinetic term.  
In the complete QGP this brings in new functions, the $\psi(q_a)$
\cite{interface1}.  Other functions could arise in the semi-QGP.  
For now, we defer these involved computations;
since the corrections $\sim g^2$ are large, $\sim 50\%$,
our results should be considered as tentative.

Besides the 't Hooft loop, which is an
interface tension for an order-order interface at $T \geq T_c$,
the interface tension for 
the order-disorder interface, at $T_c$, is also computable in
our model.  This only exists for a first order transition; for
three colors, 
\beq
\sigma_{dis} = 0.0258012 \; \frac{T_c^2}{\sqrt{g^2}} \; .
\eeq
It is necessary to compute the
corrections $\sim g^2$ before comparing to lattice data, though.

The parameters for three colors, Eq. (\ref{parameters_three_colors}),
and two colors, Eq. (\ref{parameters_two_colors}), are similar;
the difference is commensurate with a dependence on $\sim 1/N^2$,
with the coefficient of order one.  We have then assumed that
the parameters for three colors are close to those for higher $N$.
We find reasonable agreement for the interaction
measure to lattice results \cite{panero}.  
When $N \geq 4$, there is more than one 't Hooft loop.
Lattice simulations find that they obey Casimir scaling to 
good approximation 
\cite{interface_lattice2}.  We have not checked this explicitly,
but suspect that in our model, 't Hooft loops respect Casimir scaling.

The most novel prediction of our model is that there is a Higgs effect
in the semi-QGP.  This was noted first in Ref. \cite{higgs}, and
in theories at a femtoscale \cite{matrix_other}.  
To understand it, consider the quantum
fluctuations about the background field of Eq. (\ref{background}):
\beq
\left\langle \left(A_0^{qu}\right)_{a b}(\vec{x}) \, 
\left(A_0^{qu}\right)_{b a}(0) \right\rangle
\sim \int \frac{d^3p}{(2 \pi)^3} \; {\rm e}^{i \vec{p}\cdot \vec{x}}  
\sum_{n= -\infty}^{+ \infty} 
\Delta_{00}
\eeq
where $\Delta_{00}$ is the quantum propagator
\beq
\Delta_{00} =
\frac{{\rm e}^{- i p_0  \tau}}{(\vec{p}\,)^2 + p_0^2 + m_{\rm D}^2(q) }
\, ; \,
p_0 = 2 \pi T \left( n + q_a - q_b \right) \; .
\label{Higgs}
\eeq
The shift in the energy, $p_0 = 2 \pi T n \rightarrow
2 \pi T (n + q_a - q_b)$, is because we are expanding about a background
field.  The background field $A_0^{cl}$  acts upon
quantum fluctuations like an adjoint Higgs field.  Because the field
is diagonal in color, Eq. (\ref{background}), diagonal fields do not
feel the background field.  Thus for diagonal fields, the only mass they
develop is the Debye mass, $m_{\rm D}$.  This is of order $\sim g T$ times
a function of the $q_a$'s \cite{yk}.  In contrast, off diagonal fields
have non-trivial commutators with a diagonal field, and so they develop
``masses'' which are large, $\sim 2 \pi T (n + q_a - q_b)$.  

\begin{figure}[<as soon as possible>]
\begin{center}
\includegraphics[width=60mm]{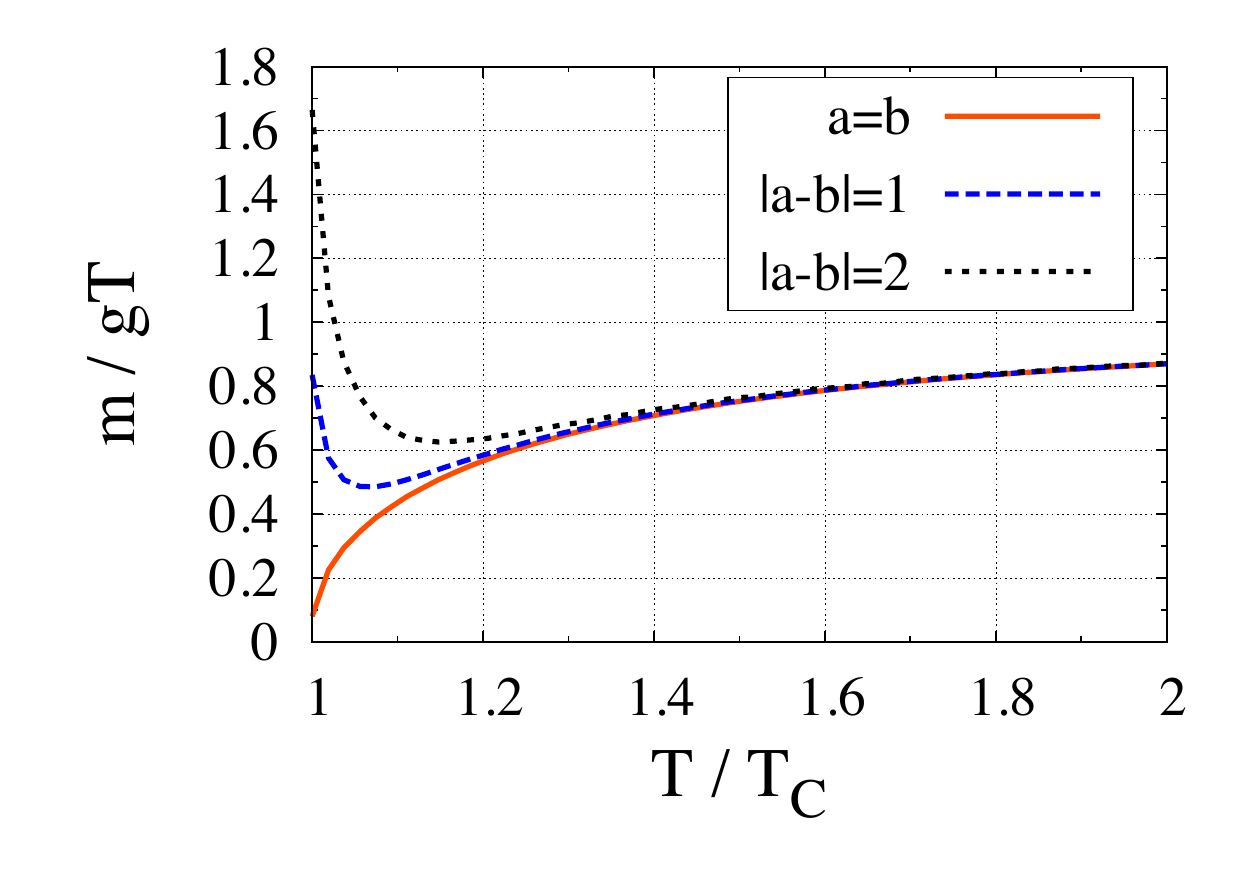}
\caption{Gluon masses, $m_{a b}/(g T)$, for 
$SU(3)$ pure gauge theory: diagonal gluons, with
$a=b$, are light; while there are two off diagonal gluons with heavy masses,
for $|a-b| = 1$ and $|a - b|=2$. }
\label{fig:3Debye}
\end{center}
\end{figure}

We illustrate this in Fig. (\ref{fig:3Debye}) for three colors.  
The masses of the two diagonal gluons are equal, and decrease as 
$T \rightarrow T_c^+$.  There are two types of off-diagonal gluons:
four with $|a-b| = 1$, and two with $|a-b|=2$.  The splitting of the masses
is evident only close to $T_c$, for $T < 1.2 \, T_c$.  

We do not plot lattice data, because it is somewhat contradictory.
Lattice measurements of a gauge invariant quantity, the two point function
between Polyakov loops, shows that the associated mass decreases
as $T \rightarrow T_c^+$  
\cite{pressure3}.  In contrast, the two point function of gluons indicate
that the gauge dependent mass increases as $T \rightarrow T_c^+$
\cite{gluon_mass_1}.  Clearly it would be best 
to reanalyze the lattice data with a Higgsed propagator in
the effective theory, 
Eq. (\ref{Higgs}), with its characteristic
combination of modes whose masses both increase and decrease.

The static, spatial gluon fields, the $A_i$, also undergo 
a Higgs effect.  This happens as well in a monopole gas \cite{shuryak}.

\begin{figure}[<as soon as possible>]
\begin{center}
\includegraphics[width=60mm]{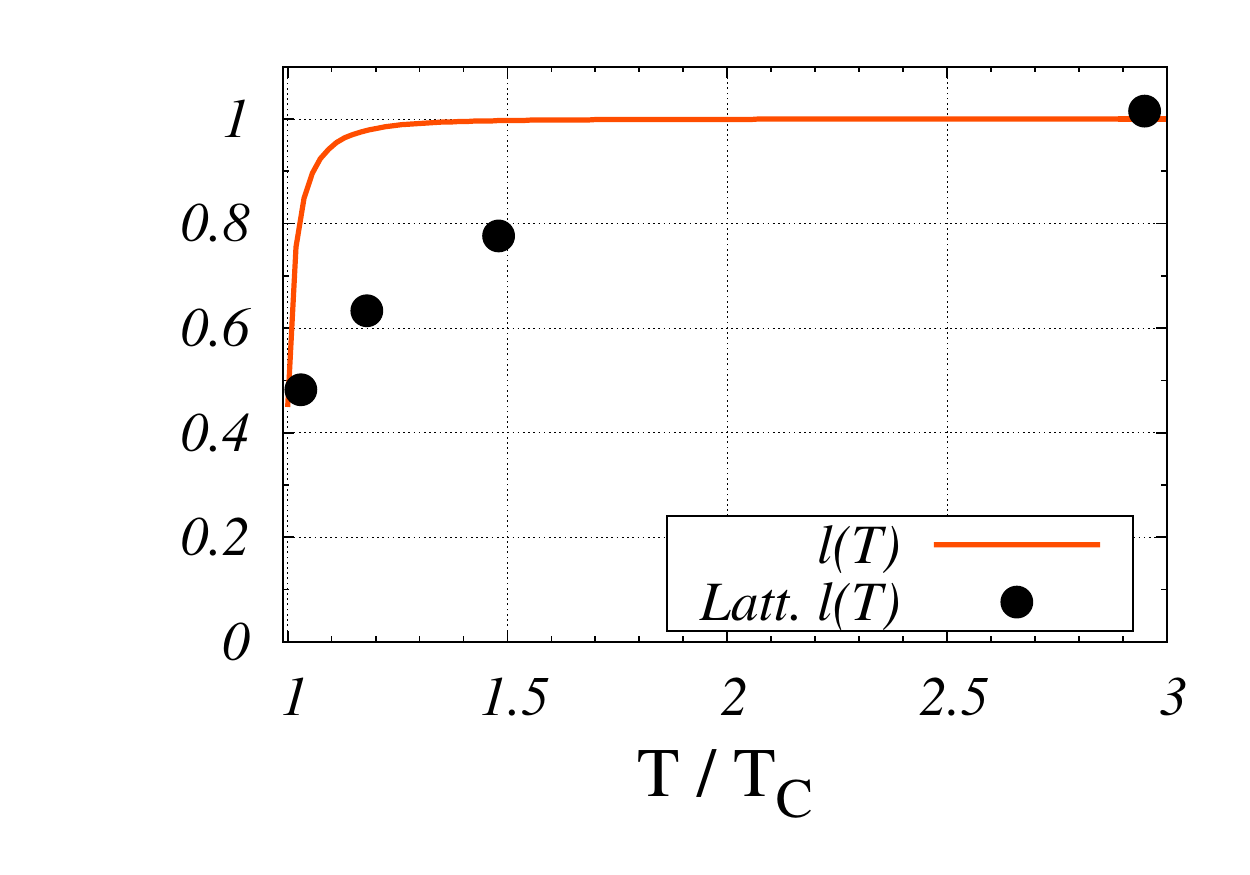}
\caption{The Polyakov loop for a $SU(3)$ pure gauge theory
from lattice simulations \cite{ren_loop3} and in our model.}
\label{fig:3Loop}
\end{center}
\end{figure}

We have not discussed the most obvious application of our model: the
computation of the Polyakov loop.  We plot this quantity, and the lattice
results, for three colors in Fig. (\ref{fig:3Loop}).
A direct comparison of the two is somewhat misleading.  We have not
computed perturbative corrections to the Polyakov loop, which enter
at $\sim g^3$ \cite{pert_ren_loop}.  This contribution is positive,
and will increase the result.  Nevertheless, while the two coincide
at $T_c$ --- which is presumably coincidence --- they immediately diverge from
one another.  From Fig. (\ref{fig:3Loop}), in our model
the loop quickly goes up to
a constant value by $\sim 1.2 \, T_c$; this is very different
from lattice measurements, for which it is not constant until 
a much higher temperature, $\sim 4.0 \, T_c$ 
\cite{ren_loop1,ren_loop2,ren_loop3}.

If our model is correct,
why does the value of the Polyakov loop, computed from our model, differ
so significantly from lattice measurements of the (renormalized) Polyakov
loop?  There is an ambiguity associated with the renormalized Polyakov loop,
from the zero point energy.   In Ref. \cite{yk} we argued that perturbatively,
the zero point energy vanishes for a straight Polyakov loop.
This argument fails for a ``smeared'' loop (see, {\it e.g.},
the appendix of Ref. \cite{ren_loop2}).
If so, then the effects of smearing must be very dramatic.

We comment that a similar
rapid growth in the Polyakov loop is found in solutions of the 
Schwinger-Dyson equations \cite{pawlowski}.  
Our results do not coincide numerically, though.

To understand our results better, consider first the limit of intermediate
temperature: say above $\sim 1.5 \, T_c$, and up to $\sim 4.0 \, T_c$.  
In this limit the $q_a$'s are small.  
As noted before, for small $q$ 
the dominant terms are $\sim c_1 T^2 q$ in the nonperturbative
potential, Eq. (\ref{non}), and that
$\sim T^4 q^2$ in the perturbative potential, Eq. (\ref{pert_pot}).
Balancing these two gives $\langle q_a \rangle \sim - c_1/\pi^2 
(T_c/T)^2$; 
for three colors, $\langle q_a \rangle \sim 0.04\,
(T_c/T)^2$.  Thus while the theory is nominally always in a ``Higgsed''
phase, as a practical matter this effect is numerically miniscule.
Further, since the loop involves the cosine of the $q_a$'s, 
asymptotically the deviation of the loop from unity is even smaller:
$1 - \ell \sim q^2 \sim (T_c/T)^4$.  

At intermediate temperatures,
what is the origin of the term $\sim T^2$ in the pressure;
or equivalently, the behavior of the interaction measure,
$\Delta/T^4 \sim 1/T^2$?
If the $q\sim 1/T^2$, for small $q$ terms $\sim T^4 q^2$
and $\sim T^2 q$ are of order one.  Then the {\it only} contribution to 
a term $\sim T^2$ in the pressure is from the $q$-{\it in}dependent terms
in the nonperturbative potential.  In our model,
this is a single term, $\sim c_3 \, T_c^2 \, T^2$.

On the other hand, for intermediate temperatures,
even if the pressure does not probe the other terms
in the potential, the 't Hooft loop does.
By measuring a 't Hooft loop, the altered boundary conditions force
the system to probe
the $q$-dependence of the potential in a nontrivial way.  
In the present model, these are determined by the coefficients
for $c_1$ and $c_2$.  (As a constant term, the value of $c_3$ doesn't 
matter.)
In this way the lattice measurements of the
't Hooft loop \cite{interface_lattice1,interface_lattice2}
are an absolutely essential constraint on our model.  

In contrast, near $T_c$ all of the parameters of the model matter,
contribute to both the behavior of the pressure 
and the value of the $q_a$'s.  
By comparison, consider the model of 
Ref. (\cite{ogilvie1}), where $c_2 = 0$.  We take
two colors, since then the expressions are algebraically simple.
From Ref. (\cite{ogilvie1}), with $q = q_1 = - q_2$,
\begin{equation}
q_{c_2 = 0}
= \frac{1}{4} \left( 1 - \sqrt{1 - \left( \frac{T_c}{T}\right)^2 } \right) \; .
\label{q_ogilvie}
\end{equation}
When $c_2 = 0$, though, the interaction measure is much
broader than indicated by the lattice data.

In the present model, with $c_1, c_2 \neq 0$,
\begin{equation}
q
= \frac{1}{4} \left( 1 - \sqrt{1 + 
\frac{2 c_1}{c_2 + (2 \pi^2/3) (T/T_c)^2} } \right) \; .
\label{q_present}
\end{equation}
The values of $c_1$ and $c_2$ are dictated by fitting the pressure, or more
sensitively, the interaction measure.  
By making $c_2$ increasingly negative from $c_2 = 0$, one finds
that the peak in the interaction measure becomes increasingly sharp.
Since the loop vanishes at $T_c$ for two colors, 
$c_1$ is then adjusted so that $q=1/4$ at $T=T_c$.

Tuning $c_2$ and $c_1$
in this way, one finds that as the peak in the interaction
measure sharpens, that the region in which $q$ is nonzero narrows
as well.  This is due to the particular values of our fit:
near $T_c$, the term $\sim c_2 T_c^2 T^2$ is not only large,
but approximately cancels the similar term,
with coefficient $(2 \pi^2 /3)\, T^4$, in the perturbative potential.
Requiring $q=1/4$ at $T=T_c$ fixes
$c_2 + 2 \pi^2/3 = - 2 c_1$, so that $|c_1|$ is small, 
$|c_1| \ll |c_2|$.  
From Eq. (\ref{q_present}), the combination 
$c_1/(c_2 + (2 \pi^2/3) (T_c/T)^2)$
enters into $q$, and implies that it is much sharper than the corresponding
factor, $(T/T_c)^2$ for $c_2 = 0$, Eq. (\ref{q_ogilvie}).
A similar cancellation happens for three colors.

We suggest that this reflects real physics: the sharpness of the interaction
measure reflects the narrowness of the region in which the loop
deviates significantly from unity.  
For the model with $c_2 = 0$, Eq. (\ref{q_ogilvie}),
$q$ always exceeds the corresponding value in our model,
with $c_2 \neq 0$.  This is also seen for $T \gg T_c$:
when $c_2 = 0$, $q \sim 0.125 \, (T_c/T)^2$;
with $c_2 \neq 0$, $q \sim 0.011 \, (T_c/T)^2$.

It is also worth contrasting our results with those in a Polyakov 
loop model.
Consider a theory which only involves the Polyakov loop of Eq. (\ref{loop}),
\beq
{\cal V}_{eff}(\ell) =
\left(- \frac{b_2}{2} |\ell|^2 
+ \frac{1}{4} ( \left|\ell\right|^2)^2 \right)
b_4 \, T^4 \; ;
\label{loop_potential}
\eeq
see, {\it e.g.}, Eq. (2) of Ref. \cite{semi2}, and 
Polyakov Nambu-Jona-Lasino (PNJL) models \cite{pnjl}.  
For three colors, the $Z(3)$ symmetry also allows a cubic term,
$\sim \ell^3 + (\ell^*)^3$, but its addition would only
complicate the algebra, and not our qualitative conclusion.
There is no cubic term for two colors.

The minimum of the potential is $\ell_0 = \sqrt{b_2}$, which
we choose to be real.  
As it is related to the pressure of an ideal gas of gluons,
we assume that the coefficient $b_4$ is independent of the temperature,
and that only $b_2$, or equivalently $\ell_0$, depends upon $T$.
The pressure and the interaction measure are then
\beq
\frac{p}{p_{ideal}} =  \ell_0^4 \;\; ; \;\;
\frac{e - 3 \, p}{4 \, p_{ideal}} \, = \;
\ell_0^3 \;T \, \frac{\partial \, \ell_0}{\partial \, T}
\; .
\label{loop_thermo}
\eeq

Consider first intermediate temperature, where the expectation value of the
loop is near one.  
To obtain a term $\sim T^2$ in the pressure,
in a loop model 
it is necessary to assume that the loop deviates from one as 
$1 - \ell_0 \sim (T_c/T)^2 + \ldots$.  This is contrast to our matrix
model, where the deviation of the loop is $\sim 1/T^4$, but there is
still a term $\sim T^2$ in the pressure.

Near $T_c$, from Eq. (\ref{loop_thermo}) a peak in the interaction measure 
corresponds to a rapid change in $\ell_0$.  This is similar to what
we find in a matrix model.  
For $\ell$ to decrease as $T \rightarrow T_c^+$, so must
$\sqrt{b_2}$ in Eq. (\ref{loop_potential}).
This is proportional to the mass of the $\ell$ field,
and so the $\ell$ mass decreases, like that of 
the diagonal modes in the matrix model, Fig. (\ref{fig:3Debye});
it is in contrast to the masses of the off-diagonal modes,
which increase.

In Ref. \cite{semi2} and other loop models \cite{pnjl}, in order
to fit the pressure the temperature dependence of the pressure,
$b_2$ must have a complicated form.  In
our matrix model, the coefficients are
just $\sim T^4$, Eq. (\ref{pert_pot}), and 
$\sim T^2$, Eq. (\ref{non}).
In a mean field theory such as this, simplicity is a virtue.

Lastly, the splitting of gluon masses near
$T_c$ is special to a matrix model, as a Higgs effect for the adjoint
scalar $A_0$ field.  See, {\it e.g.}, the loop model
of Ref. \cite{ogilvie2}, where the splitting of masses does not occur.

Our analysis is a preliminary first step.  
In deriving our results, we balance the perturbative potential,
${\cal V}_{pt}(q_a)$, against the non-perturbative potential,
${\cal V}_{non}(q_a)$.  In powers of $g^2$, the perturbative
potential is of order one, so implicitly we have assumed that
the non-perturbative is as well.  Since the non-perturbative
potential represents a resummation of effects to all orders,
this is a strong assumption.  Nevertheless, it allows us to envisage
computing to higher order in $g^2$.  
Corrections at least to order $\sim g^2$
and $\sim g^3$ are needed in order to make a serious comparison to lattice
data.  This also requires precise lattice data, close to the continuum
limit, not just for the pressure, but also for the 't Hooft loop and
gluon masses.  

There are several formal questions raised by our analysis.  
The parameters of effective theories can be 
computed from lattice 
simulations \cite{lattice_eff};
doing so for elements of the Lie algebra, 
instead of for elements of the Lie group, may
be useful.  It is also
necessary to extend the analysis of Hard Thermal Loops in the complete
QGP to the semi-QGP.  This is equivalent to
understanding the 
analytic continuation of the thermal Wilson line from imaginary to real time.  

To compare with QCD it is necessary to include the
effects of dynamical quarks. 
It will be especially interesting 
to see if, upon adding the
effects of quarks to the perturbative potential, ${\cal V}_{pt}(q_a)$,
whether the thermodynamics \cite{pressure_dyn}, 
and the Debye mass, are reproduced using
the same parameters for the non-perturbative potential,
${\cal V}_{non}(q_a)$, in the pure glue theory.  
(With dynamical quarks, the 't Hooft
loop does not exist as an order parameter.)

Without detailed computation, we assume 
that a narrow width
for the semi-QGP in the pure glue theory implies the same for QCD.
We thus conclude with some speculations for the phenomenology of heavy ion
collisions.

If RHIC probes to some temperature in the QGP, then LHC may probe
to a temperature approximately $\sim 50\%$ higher.  
If the AdS/CFT correspondence
holds for QCD, then results at the LHC must mimic those at RHIC.  With
the present analysis, the picture is rather more complicated.

We assume, for the sake of argument, that RHIC probes only to a temperature
in the semi-QGP, very near $T_c$.  Then the LHC begins at a temperature
well in the complete QGP.  Any conclusions are tempered by the fact that
even if the LHC starts at a higher temperature, as it cools it must traverse
through the semi-QGP.

In the semi-QGP, the ratio of $\eta/s$
decreases as the square of the Polyakov
loop as $T \rightarrow T_c^+$; this is true both in the
pure glue theory, and with dynamical quarks \cite{yk}.  
Conversely, then, $\eta/s$ increases as the temperature goes 
up from $T_c$.  This is in sharp contrast to models
based upon the AdS/CFT correspondence, where
$\eta/s$ is constant \cite{gubser,kiritsis,other}.
Unfortunately, a computation beyond leading logarithmic order is required
to compute the precise dependence of $\eta/s$ with temperature.

If the shear viscosity increases strongly from $T_c$, {\it and} the
system is in thermal equilibrium, then an increased shear viscosity should
lead to an increase in particle multiplicity, and a decrease in the elliptical
flow, over the results expected from a (nearly) ideal gas.  If the
shear viscosity increases significantly, though, a hydrodynamic description
could easily break down.  

It is also possible that the temperature dependence of $\eta/s$
is weak; if so, then
the particle multiplicity and elliptical flow at LHC should be similar
to that expected by an extrapolation from the results at RHIC.  There
are then other ways to probe the effects of the semi-QGP.

Consider, for example, energy loss, which is 
controlled by a parameter $\hat{q}$.  In
the complete QGP, $\hat{q} \sim T^3$, 
or equivalently, the entropy density, $s$.
In kinetic theory, $\hat{q}/s$ and $s/\eta$ are each proportional
to a cross section, so one expects that a minimum in $\eta/s$
corresponds to a maximum in the energy loss, $\hat{q}/s$
\cite{shuryak,energy_loss}.
Following the methods of Ref. \cite{yk}, 
the energy loss of a quark can be computed in the semi-QGP;
as $T \rightarrow T_c^+$, it vanishes linearly in the
Polyakov loop.
Thus in the semi-QGP, both $\eta/s$ and $\hat{q}$ decrease
as $T \rightarrow T_c^+$; the difference from Refs. \cite{shuryak,energy_loss}
is because the kinetic theory for 
the semi-QGP is in the presence of a background $A_0$
field.  As the
temperature increases from $T_c$, then,
excluding the obvious dependence upon 
the entropy, the energy loss is larger in the
complete QGP than in the semi-QGP.  
As with the shear viscosity, determining the precise
dependence upon temperature 
requires computation beyond leading logarithmic order.  

There is also a qualitatively new phenomenon in the semi-QGP: besides
energy loss, the propagation of a colored field
is suppressed by the background $A_0$ field \cite{yk}.  This suppression
is universal, independent of the mass or momentum of the colored field.
A complete analysis need incorporate this universal suppression
as well as energy loss.

Lastly, we note that given the modified propagator of the semi-QGP, 
Eq. (\ref{Higgs}), there
are also significant modifications to the heavy quark potential
\cite{dumitru_kyoto}.  This
can also be compared to lattice data, which we defer for now.

In the end, our speculations will soon be rendered moot by the wealth
of results which will flow from heavy ion collisions at the LHC.  The present
approach is based upon constructing an effective theory from 
the results of lattice
simulations; not just of the pressure, but quantities such
as the 't Hooft loop and screening masses.
This can then be tested against 
predictions from the AdS/CFT correspondence 
\cite{ads_reviews,gubser,kiritsis,other,gyulassy} and other models
\cite{shuryak}.

\acknowledgements
We thank A. Bazavov, P. de Forcrand, O. Kaczmarek, F. Karsch, J. Liao, 
M. Pepe, P. Petreczky, and E. Shuryak for discussions and comments.
The research of A.D. was supported by the U.S. Department of
Energy under contract \#DE-FG02-09ER41620, and
by PSC-CUNY research grant 63382-00~41;
of Y.G., in part by the 
Natural Sciences and Engineering Research Council of Canada;
of Y.H., by the Grant-in-Aid for the Global COE
Program ``The Next Generation of Physics, Spun from Universality and
Emergence'' from the Ministry of Education, Culture, Sports, Science
and Technology (MEXT) of Japan; of R.D.P.,
by the U.S. Department of Energy under contract \#DE-AC02-98CH10886;
R.D.P.\ also thanks the Alexander von Humboldt Foundation for their
support.

\end{document}